\documentclass[12pt]{article}

\newcommand{\m}{\mathrm}
\newcommand{\be}{\begin{equation}}
\newcommand{\ee}{\end{equation}}
\newcommand{\ba}{\begin{eqnarray}}
\newcommand{\ea}{\end{eqnarray}}

\usepackage{graphicx}
\usepackage{amssymb}
\usepackage{amsmath}
\usepackage[T1]{fontenc} 
\usepackage[ansinew]{inputenc} 
\usepackage[nosort]{cite}
\newcommand{\inbar}{\vrule height1.57ex width.4pt depth0pt}
\newcommand{\SW}{\relax{\hbox{$\ \inbar\kern-.285em{\rm S}$}}}

\begin{document}
\thispagestyle{empty}
\begin{center}

\null \vskip-1truecm \vskip2truecm

{\Large{\bf \textsf{Superradiance in the Bulk Protects Quantum State Evolution of Rapidly Rotating Matter on the Boundary}}}

{\large{\bf \textsf{}}}

{\large{\bf \textsf{}}}

\vskip1truecm

{\large \textsf{Brett McInnes}}

\vskip1truecm

\textsf{\\  National
  University of Singapore}

\textsf{email: matmcinn@nus.edu.sg}\\

\end{center}
\vskip1truecm \centerline{\textsf{ABSTRACT}} \baselineskip=15pt
\medskip

It has been argued that the rate at which the interior of an AdS black hole evolves is dual to the rate of evolution of the (quantum state of the) strongly coupled matter on the boundary which, according to holography, is dual to the black hole. However, we have shown elsewhere that it seems to be possible, by adjusting the specific angular momentum of an AdS$_5$-Kerr black hole, to reduce this rate to (effectively) zero. We argue that this is unphysical, and that it is prevented by the intervention of a superradiant instability, which causes the black hole to shed angular momentum when the angular velocity exceeds a certain critical value. The precise way in which this works has recently been explained by the ``grey galaxy'' model of the end state, in which the angular momentum is transferred to a ``galactic disc.'' Thus, the black hole itself cannot sustain a specific angular momentum beyond a critical value: there is an effective upper bound. The holographic interpretation is that, beyond a certain limiting specific angular momentum, strongly coupled matter (corresponding to the black hole) will spontaneously shed angular momentum to some other, confined, form of matter (corresponding to the disc). This idea is supported by recent numerical work on ultra-vortical plasmas. Such an upper bound on specific angular momentum would prevent arbitrarily small rates of quantum state evolution on the boundary. We give a tentative discussion of the relevant observational data in the case of the vortical Quark-Gluon Plasma, and suggest a way in which such an upper bound might appear in future observations.

\newpage

\addtocounter{section}{1}
\section* {\large{\textsf{1. Rotation and the Evolution of the Quantum State}}}
Entering a Schwarzschild black hole is ill-advised, \emph{not} because one is ``inevitably pulled into the central singularity'', but rather because \emph{the interior geometry is dangerously time-dependent}. The singularity is spacelike and lies to the future: it cannot harm anyone. What actually happens is that the rules of spatial geometry inside the event horizon change as time passes, and one finds that the ambient space becomes progressively less habitable. That is, the dynamic spatial geometry requires the shape of any intruder to \emph{change}, in some disagreeable manner. The geometry evolves in such a way that, in fact, that the extent of this change is unbounded. This is the correct formulation of the so-called singularity ``problem''.

The fact that the interior is a region of ceaseless change\footnote{Like the Schwarzschild black hole, the AdS$_5$-Kerr black hole we study here has no timelike Killing field anywhere in its interior. It seems likely that all black holes share this property when perturbations of inner horizons are taken into account: see particularly \cite{kn:tyler}.} is its characteristic feature, distinguishing it sharply from the typical black hole exterior. For example, the fact that Einstein-Rosen bridges (internal ``wormholes'') necessarily collapse rapidly is responsible for the fact that, classically, they cannot be traversed \cite{kn:viss}. This is directly relevant to current efforts to construct (quantum) traversable wormholes \cite{kn:lobo,kn:aronwall,kn:juan,kn:bilotta}.

This has serious consequences when one attempts to extend the bulk-boundary duality \cite{kn:nat} to the interior of asymptotically AdS$_5$ black holes\footnote{Throughout this work, the bulk is five-dimensional.}. We assume that the exterior of the black hole has attained equilibrium with its Hawking radiation, which is possible \cite{kn:ruong} for an AdS$_5$ black hole which is ``large'' in the technical sense. This exterior is described dually by the thermal equilibrium state of the boundary matter. But now we must ask: how can a static boundary state possibly account for the interior of the black hole, when the latter is anything but static? This basic question has attracted much attention of late, in connection with ``complexity'' \cite{kn:underwood}: see \cite{kn:poli} for a survey.

In short: if holography is valid everywhere in an AdS black hole spacetime, something must be changing on the boundary, at a certain rate. The task is to understand this ``something'' or at least its rate of change.

The key observation here is that there is a crucial difference \cite{kn:scott} between the nature of ``thermal equilibrium'' as it is understood classically and its meaning for a (closed) quantum system. For in the latter case, the evolution of the quantum state of the boundary matter \emph{continues unabated}, even after ordinary thermal equilibrium has been attained; it only ceases after an enormously longer time. It is very tempting to propose that \emph{this essentially unending boundary quantum state evolution is holographically dual to the never-ending evolution of the bulk black hole interior}, and a particular interpretation of this proposal is the subject of the works surveyed in \cite{kn:poli}. Some new evidence for the existence of this kind of duality was presented in \cite{kn:111}.

The black hole interior is controlled, just as the exterior is, by the black hole parameters, such as its Hawking temperature and specific angular momentum; and it this simple yet essential observation that opens the way to investigating the physics of the proposed ``extended duality''. For it follows that varying these parameters must influence the rate at which the quantum state evolution on the boundary proceeds.

Now it was shown in \cite{kn:111} that, in the absence of any other effect, \emph{it is possible in this way to drive this rate down essentially to zero}. This is done by taking the specific angular momentum of the black hole to be sufficiently large, while fixing its temperature; it is imposed as soon as thermal equilibrium is established.

But this does not make sense: it means that, for a rapidly rotating AdS black hole, the boundary Hamiltonian is somehow ``nullified'', for any given temperature, however high. This absurdity implies that, for the extended duality to be physical, some kind of \emph{upper bound on the specific angular momentum} of the black hole must exist, given its temperature. By an application of ``ordinary'' holographic duality, it follows that such a temperature-dependent bound must exist for the specific angular momentum of strongly coupled matter.

The obvious move is to search for some kind of instability triggered by high specific angular momenta in actual examples of strongly coupled matter, particularly those provided by collisions of heavy ions. The existence of instabilities in such matter at high vorticities \emph{has in fact been predicted}, by numerical simulations: see particularly \cite{kn:chern3} (also \cite{kn:chern1,kn:chern2,kn:chern4}).

The problem now is to identify the dual phenomenon in the bulk. We propose that the dual is a \emph{superradiant} instability \cite{kn:super} of the bulk black hole. We will also propose that, with the aid of the recent identification of the superradiant end state for these black holes in terms of ``grey galaxies'' \cite{kn:grey}, we may be able to predict qualitatively what forthcoming experimental data will reveal about the specific angular momenta of ultra-vortical plasmas.

We begin with a very brief r\'{e}sum\'{e} of the duality.

\addtocounter{section}{1}
\section* {\large{\textsf{2. The Interior Evolution / Boundary State Evolution Duality.}}}
Unfortunately it is not known exactly how to measure the rate of evolution of the quantum state of strongly coupled matter. It can however be argued on very general grounds (see \cite{kn:jacob} for a clear explanation) that, in the simplest cases, the rate is proportional to the temperature of the strongly coupled matter when it is formed, and to the total entropy of the system then. (See \cite{kn:tallarita} for more details regarding the limitations of this proposal.)

The total entropy of the system is however not a practical parameter for describing actual examples of strongly coupled matter: worse, its rate of change is not a thermodynamically intensive quantity, as the rate of evolution of the quantum state should be.

For this reason, and also guided by holographic considerations (see below), we proposed in \cite{kn:111} that the correct parameter to use here is the \emph{specific} entropy, the entropy per unit mass\footnote{In practice: the ratio of the entropy density to the mass density.}. The idea is that what counts here is not the total entropy but rather \emph{the entropy contributed by each particle}, and it is \emph{this} quantity that influences the ``rate of evolution on the boundary''.

To be precise, then, the rate we seek to study is given simply by
\begin{equation}\label{ALPHA}
\mathfrak{R} \;=\; {\mathfrak{s}\,T\over \hbar},
\end{equation}
where $T$ is the temperature, $\mathfrak{s}$ is the specific entropy, and $\hbar$ is the reduced Planck constant; see \cite{kn:111} for more detail\footnote{The units of $\mathfrak{R}$ are of the form 1/(mass $\times$ time); actual values are given as multiples of $c^2/\hbar,$ so $\mathfrak{R}$ is dimensionless in natural units.}. (Gothic letters denote intensive quantities throughout this work.)

It is natural to ask how $\mathfrak{s}\,T/\hbar$ behaves when evaluated on actual strongly coupled matter, such as the \emph{Quark-Gluon Plasma} (QGP) \cite{kn:STAR,kn:franc,kn:prefut,kn:hot,kn:sign,kn:overview}. The (by far) best understood examples of the QGP are produced in \emph{central} (``head-on'') collisions of heavy ions at facilities such as the RHIC and LHC. For central collisions, ingenious techniques have been developed \cite{kn:olli,kn:soundspeed} which allow experimental determinations of entropy densities, temperatures, and other parameters of the QGP; and sophisticated phenomenological models of these ``central plasmas'' also exist (see for example \cite{kn:sahoo}).

One finds in this way that $\mathfrak{s}\,T/\hbar$ is a monotonically increasing function of temperature, \emph{but this function is bounded both below and (asymptotically) above}. Thus, in this case, we claim that there are bounds below and above on the rate at which the quantum state evolution on the boundary proceeds. In fact it turns out that the rate only varies over a rather narrow (relative) range, between about $0.9 \times c^2/\hbar$ and $1.3 \times c^2/\hbar$.

In \cite{kn:111} we asked whether a holographic model, in which (as above) $\mathfrak{s}$ and $T$ are obtained by studying the entropy per unit mass and the Hawking temperature of a ``large'' AdS$_5$-Schwarzschild black hole, could replicate these findings regarding $\mathfrak{s}\,T/\hbar$. We found that it could; this will be summarised briefly in the next section. The success of holography in this simple case encourages us to believe that it might work in more general cases.

Perhaps the most important such case is the locally rotating (``vortical'') version of the QGP \cite{kn:STARcoll,kn:jiang,kn:becca}, which is far harder to probe either phenomenologically or experimentally. These plasmas are produced in collisions which are \emph{not} central, so that the special techniques for handling central collisions \cite{kn:olli,kn:soundspeed} cannot be used. This is a case where holography might well prove to be useful.

The bulk is taken to be a singly-rotating  AdS$_5$-Kerr black hole. Now $\mathfrak{s}\,T/\hbar$ is a function of two variables, temperature and \emph{specific angular momentum}\footnote{The specific angular momentum of the boundary matter is the ratio of the angular momentum density to the mass density. In the dual bulk, it is the ratio of the black hole angular momentum to its mass.}. The two main findings of \cite{kn:111} in this case were as follows.

$\bullet$ There is still a universal least upper bound on $\mathfrak{s}\,T/\hbar$ for all temperatures and specific angular momenta, but this upper bound is somewhat higher than in the non-rotating case. Thus, the model makes a definite prediction: for some ranges of temperatures and specific angular momenta of the boundary matter, it should be possible to observe values of $\mathfrak{s}\,T/\hbar$, in strongly coupled matter, which are \emph{strictly forbidden} in the non-vortical case.

$\bullet$ Surprisingly, the universal upper bound on $\mathfrak{s}\,T/\hbar$ occurs at high \emph{but not extremely high} specific angular momenta. That is, at each given temperature, $\mathfrak{s}\,T/\hbar$ increases at first with increasing specific angular momentum; but eventually it reaches a maximum, and then decreases to \emph{arbitrarily small values}, including, once again, values forbidden to non-vortical plasmas.

The key point here is that we have an interesting example of what might be called ``\emph{rotational holography}'' making (potential) contact with observations: the theory predicts that sufficiently vortical plasmas should give rise to values of $\mathfrak{s}\,T/\hbar$ both above and below values attainable in non-vortical plasmas. (On the other hand, if even higher values, beyond the predicted maximum, should be observed, then the theory will be falsified.) As we will see later, the predicted maximum is not much higher than values that have already been deduced from experiments, so this may well be a realistic possibility. If ``rotational holography'' were to be confirmed in this way, one would certainly be inclined to take seriously its claims regarding less readily observed phenomena, such as post-equilibrium quantum state evolution on the boundary.

However, these results have a less welcome implication for the proposed extension of holography to the black hole interior: as was mentioned in the previous Section, the fact that the model predicts arbitrarily low values for $\mathfrak{s}\,T/\hbar$ means that the quantum state evolution of the boundary matter effectively does not proceed at all in the presence of very large vorticities.

Note that very extreme vorticities, and their potentially observable consequences for the structure of the QGP, are currently being actively studied, both theoretically \cite{kn:buzztuch} and experimentally \cite{kn:newalice}. That is, it is conceivable that extremely large vorticities might be experimentally accessible at some point. Note also that the prediction of arbitrarily low values for $\mathfrak{s}\,T/\hbar$ is not some peculiarity associated with near-extremal black holes (which have very low Hawking temperatures): as mentioned earlier, the effect is predicted to occur at \emph{any} temperature.

There is in fact a classical effect which \emph{might} prevent arbitrarily low values for $\mathfrak{s}\,T/\hbar$ in the AdS bulk: black hole \emph{superradiance} \cite{kn:super}. This is a potential instability of rotating and charged black hole spacetimes in which there is a mechanism for ``trapping'' Hawking radiation, as of course there is for ``large'' AdS black holes. This phenomenon is particularly interesting because it has been suggested that it might actually occur, under some conditions, in astrophysical black holes: see for example \cite{kn:aoguo,kn:silk}.

Until recently, however, it has been difficult to exploit superradiance as a way of restricting asymptotically AdS black hole spacetimes, simply because the end state of the instability in the AdS case was not clearly understood. Now, however, this problem has been explored in great detail in \cite{kn:grey}. There it is found that the end state is a ``\emph{grey galaxy}'' consisting of a black hole \emph{rotating at a critical angular velocity} (corresponding to the onset of superradiance), surrounded by a flat disk of gas which turns out \cite{kn:tidal} to be weakly coupled to the black hole. The flat disc is of course radically different, geometrically and physically, to the black hole. In short, the black hole bifurcates into two distinct systems, shedding some angular momentum to the disc; it cannot rotate steadily at arbitrarily high angular momenta.

The authors of \cite{kn:grey} propose the following schematic holographic interpretation of this result: that strongly coupled matter is destabilised by ultra-rapid rotation, shedding angular momentum to some other form of matter (which does \emph{not} have a holographic interpretation in terms of an AdS black hole)\footnote{In \cite{kn:grey} this is formulated as follows: ``the black hole part of the Grey Galaxy solution is a ``quark-gluon plasma''; an interacting configuration of $N^2$ ``gluons''. On the other hand, the gas part of the Grey Galaxy solution is composed of local singlets, and so can be thought of as the moral analogue of ``glue-balls''. In field theory terms, as (the angular velocity approaches the critical value), the quark-gluon plasma manages to increase its entropy by expelling fast-rotating glue-balls $\ldots$''}. The strongly coupled matter then rotates more slowly, going back to being on the brink of instability, corresponding to the superradiance condition being saturated by the bulk black hole. This effectively imposes an upper bound on the specific angular momentum of the black hole, and of the system dual to it: the surviving strongly coupled matter.

Notice that we are \emph{not} claiming that there is an upper bound on the specific angular momentum of the \emph{whole system}, and indeed in \cite{kn:grey} there is no such restriction. The bound is only on the specific angular momentum of the black hole (and on that of the part of the dual system corresponding to it).

Remarkably, this picture is supported by very recent numerical modelling \cite{kn:chern3} which suggests that, at very high specific angular momenta, the vortical QGP becomes unstable, in the sense of breaking up into an inhomogeneous system which is partly deconfined and partly confined. This spontaneous bifurcation in the phase space is modelled holographically by the bifurcation of the spinning AdS black hole into the components of the grey galaxy, with \emph{only} the core black hole modelling the surviving deconfined plasma.

This picture is clearly far too simple. But it might be a useful first step towards a more nearly adequate holographic account of the vortical plasma. Let us see if it can be made to work, even at this basic level.

We strongly emphasise that it is far from clear that superradiance will play the role we need it to in this case. Firstly, it is not clear that superradiance occurs at all for these black holes: in \cite{kn:109} we used the customary relation of the geometric mass parameter to the physical mass, and we found that the AdS$_5$-Kerr black hole never experiences superradiance, at any specific angular momentum. Here however we will use a corrected formula, given in \cite{kn:gaogao}, and we will see that this changes the situation: superradiance does indeed occur. The point is that this was not inevitable.

Secondly, we need superradiance to keep $\mathfrak{s}\,T/\hbar$ well away from zero, \emph{independently of the temperature}. This is not easy to achieve.

Here we will show that superradiance \emph{imposes a universal lower bound on $\mathfrak{s}\,T/\hbar$} for these black holes: the lower bound is approximately $0.80 \times c^2/\hbar,$ which in fact is quite a high value by comparison with values for ordinary matter. Thus we conclude finally that it is not possible in this manner to halt quantum state evolution in strongly coupled matter, or even to come close to doing so.

Let us discuss the details. We begin with the simplest, non-rotating case.

\addtocounter{section}{1}
\section* {\large{\textsf{3. The AdS$_5$-Schwarzschild Case.}}}
We begin by considering $\mathfrak{s}\,T/\hbar$ for an actual physical example of strongly coupled matter, the Quark-Gluon Plasma (QGP) produced in heavy-ion collisions \cite{kn:STAR,kn:franc,kn:prefut,kn:hot,kn:sign,kn:overview}. Very recent advances \cite{kn:olli,kn:soundspeed} in the study of the QGP do in fact permit experimental estimates of $\mathfrak{s}$ and $T$ in the case of \emph{central} (``head-on'') collisions: the entropy density is determined by measuring the number of emitted charged particles in the final state (the ``multiplicity'') and the temperature is found from the transverse momentum. Phenomenological models of these central collisions have also been developed, see for example \cite{kn:sahoo}.

One finds in this way (see \cite{kn:111} for a discussion) that $\mathfrak{s}\,T/\hbar$ behaves in a straightforward way as the impact energy of the central collision (therefore, the temperature of the plasma) is increased: it rises monotonically from a minimum value towards an asymptotic upper bound. The lower bound (roughly $0.9 \times c^2/\hbar$) is readily understood: the QGP simply cannot exist at ``low'' temperatures (and low baryonic chemical potential): it hadronizes. The asymptotic upper bound (on the order of $1.3 \times c^2/\hbar$) is explained by simple thermodynamic considerations: the thermodynamic Euler relation in this case gives
\begin{equation}\label{A}
\mathfrak{s}\,T/\hbar\;=\; \left(1 + {p\over \varrho c^2}\right){c^2\over \hbar},
\end{equation}
where $p$ is the pressure and $\varrho$ is the mass density\footnote{We use mass rather than the more conventional energy densities for later convenience, when we come to discuss black holes.}. For a relativistic fluid (which the plasma increasingly approximates as the impact energy increases, see \cite{kn:sign}) one has $p/(\varrho c^2) \approx 1/3$ and so $\mathfrak{s}\,T/\hbar$ approaches ${4\over 3}{c^2\over \hbar}$ from below. In short, $\mathfrak{s}\,T/\hbar$, for these ``central'' plasmas, lies in a narrow range between about $0.9 \times c^2/\hbar$ and just below about $1.33 \times c^2/\hbar.$

As an aside, note that these values are very much larger than those obtained by evaluating $\mathfrak{s}\,T/\hbar$ for ordinary matter. For example, the value \cite{kn:nist} for
water at 300 K is about 11 orders of magnitude smaller. In simple language: the quantum state of the QGP evolves, under normal circumstances, extremely rapidly. This is of course as one would expect.

Notice that both the lower and the upper bounds probe fundamental aspects of the physics of the QGP: hadronization in one case, the approach to being a relativistic fluid on the other. Any observation suggesting values outside these bounds would demonstrate radically new behaviour.

If indeed it is true that $\mathfrak{s}\,T/\hbar$ measures the rate at which post-equilibrium quantum state evolution proceeds, then we conclude that, in the case of the ``central'' QGP, that rate can be varied by changing the temperature (impact energy). It varies only a little, however, and certainly cannot be reduced below a certain (large) value.

Now let us discuss how all this appears from the point of view of the holographic dual, which involves an asymptotically AdS$_5$-Schwarzschild black hole. As this is just a special case of the AdS$_5$-Kerr case to be discussed below, we will be brief.

Before we begin, we wish to stress that $\mathfrak{s}\,T/\hbar$ is a quantity particularly well-behaved from a holographic point of view. The bulk black hole is described by quantities such as the black hole entropy, which depends on the AdS$_5$ Newton constant $G_5$. We wish to avoid such dependence in a holographic treatment, because the actual value of $G_5$ is not known. (It can of course be expressed in terms of quantities on the boundary \cite{kn:nat}, but the numerical values of these are also unknown.) Now $\mathfrak{s}$ and $T$ correspond holographically to the black hole entropy \emph{per unit mass} and to the Hawking temperature, and neither of these has any explicit dependence on $G_5$ (see below); so they are natural parameters when we switch back and forth between boundary and bulk.

The temperature is expressed in the familiar way in terms of $L$, the asymptotic AdS$_5$ curvature length scale, and of $r_{\textsf{H}},$ the value of the radial coordinate at the event horizon. Inverting that relation, we regard $r_{\textsf{H}}$ as a function of the temperature, and this allows us to regard $\mathfrak{s}\,T/\hbar$ as a function of $T$. It is convenient to define $\Gamma(T)$ as the dimensionless version of this function (that is, $\Gamma$ is $\mathfrak{s}\,T/\hbar$ expressed as multiples of $c^2/\hbar$), and to set $T^* = k_{\textsf{B}}T/(\hbar c),$ where $k_{\textsf{B}}$ is the Boltzmann constant, so that $T^*L$ is dimensionless.

One finds (see \cite{kn:111} for the details) that the graph of $\Gamma$ is as shown in Figure 1.

\begin{figure}[!h]
\centering
\includegraphics[width=0.80\textwidth]{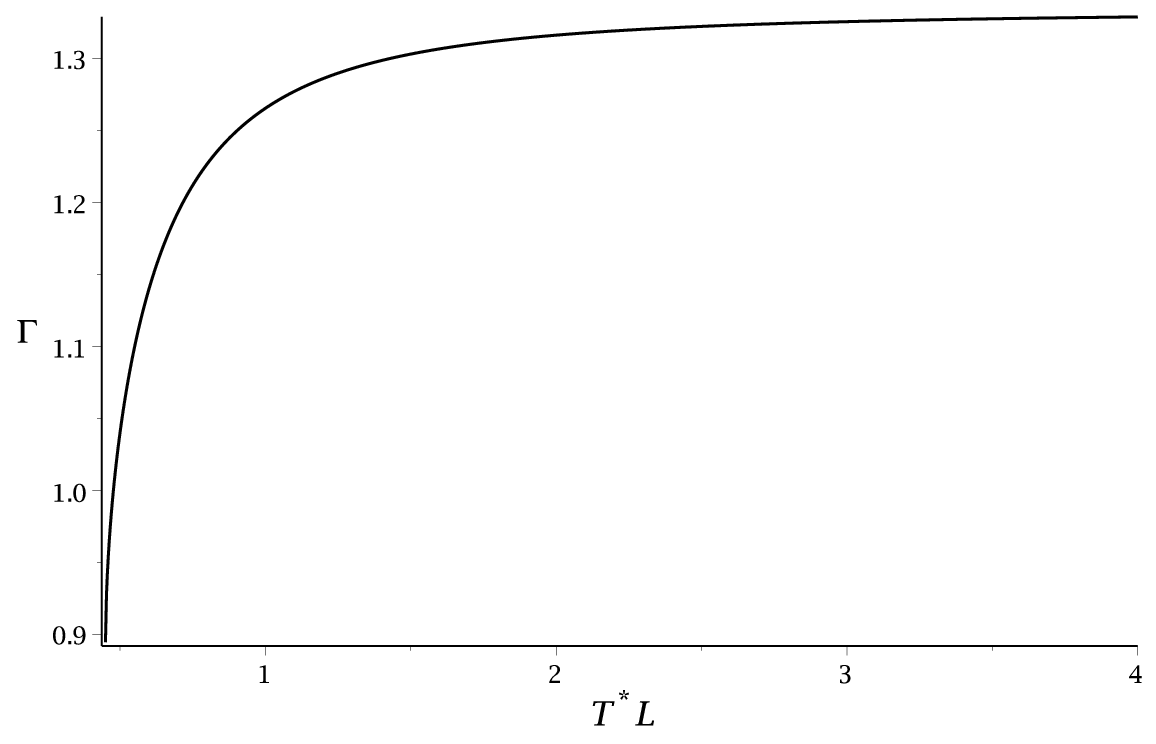}
\caption{$\Gamma$, the dimensionless version of $\mathfrak{s}\,T/\hbar$, for AdS$_5$-Schwarzschild, as a function of the dimensionless temperature $T^*L.$}
\end{figure}

We see that the holographic model predicts that $\Gamma$ has a minimum value at the smallest possible value of the temperature. It is well known \cite{kn:edwit} that such black holes do indeed have a minimum possible temperature: the Hawking temperature cannot be smaller than
\begin{equation}\label{B}
T_0 \;=\; \sqrt{2}\hbar c / \left(\pi k_{\textsf{B}}L\right).
\end{equation}
This minimum temperature corresponds holographically to the minimal possible temperature of the QGP (produced in central collisions), thought to be (for plasmas with small baryonic chemical potentials, as is the case in the plasmas generated in RHIC and LHC heavy ion collisions) around 150 MeV/$k_{\textsf{B}}$. (We do not claim of course that this simple holographic model gives an adequate account of this minimal temperature, which is in fact a smooth crossover and not a first-order phase change as it is here. We adopt the identification of $T_0$ with the crossover temperature with this warning always borne in mind.)

The value of $\Gamma$ at the minimal temperature is about $0.9.$ As the temperature increases, $\Gamma$ rises but is clearly bounded above. It can be shown \cite{kn:111} that this upper bound is precisely $4/3$. All this is in surprisingly good agreement with our earlier discussion; the appearance of the number $4/3$ is particularly striking, since the two derivations could hardly be more different.

The initial temperature of the QGP is still not known with any certainty, though progress has been reported recently \cite{kn:startemp}. Probably it does not exceed thrice the minimal temperature in even the most energetic current collisions. Now from the definition of $T^*$ we have
\begin{equation}\label{C}
T^*L \;=\; {\sqrt{2}\over \pi}\times {T\over T_0},
\end{equation}
and so we see that, in Figure 1, the range of values of $T^*L$ currently accessible to experiment is from about $0.45$ up to about $1.35$. This corresponds to a range of values for $\Gamma$ between about $0.89$ and $1.30$, again in remarkably good agreement with the data.

To summarize overall: if we model the rate of quantum state evolution in non-vortical post-equilibrium strongly coupled matter by means of the quantity $\mathfrak{s}\,T/\hbar$, then that rate is never small, and is only expected to vary with temperature through a small range. If we model the situation by using a non-rotating AdS$_5$ black hole, then we can reproduce these results very well. We should stress here that we are using a very basic holographic model. Far more elaborate models now exist (see for example \cite{kn:fancy}), and it would be of considerable interest to repeat our calculations using such a model.

This suggests that we attempt to probe the case of the vortical QGP, for which there are far fewer data, using holography. We now proceed to that.

\addtocounter{section}{1}
\section* {\large{\textsf{4. $\mathfrak{s}\,T/\hbar$ in the Vortical Case}}}
The two principal parameters to consider when discussing heavy ion collisions are the impact energy and the centrality, a measure of the extent to which the collision is not ``head-on''; collisions with non-zero centrality are studied by examining the production of $\Lambda$/$\overline{\Lambda}$ hyperons. The corresponding theoretical parameters describing the resulting QGP are the (initial) temperature of the plasma and its angular momentum density. The latter, manifested as ``vorticity'', is the subject of intense interest \cite{kn:STARcoll,kn:jiang,kn:becca}. Even very extreme cases, such that the angular velocity, $\omega$, is so large that $c/\omega$ is smaller than the mean free path in the plasma, have been studied \cite{kn:buzztuch}.

In holography, the temperature is modelled, as we have seen earlier, by using an asymptotically AdS$_5$ black hole. To model the angular momentum, we clearly need that black hole to rotate.

Proceeding as simply as possible, we will study ``large'', singly-rotating, \emph{non}-extremal AdS$_5$-Kerr black holes \cite{kn:hawk,kn:cognola,kn:gibperry,kn:gaogao}, with metric
\begin{flalign}\label{D}
g(\textsf{AdSK}_5)\; = \; &- {\Delta_r \over \rho^2}\left[\,\m{d}t \; - \; {a \over \Xi}\,\m{sin}^2\theta \,\m{d}\phi\right]^2\;+\;{\rho^2 \over \Delta_r}\m{d}r^2\;+\;{\rho^2 \over \Delta_{\theta}}\m{d}\theta^2 \\ \notag \,\,\,\,&+\;{\m{sin}^2\theta \,\Delta_{\theta} \over \rho^2}\left[a\,\m{d}t \; - \;{r^2\,+\,a^2 \over \Xi}\,\m{d}\phi\right]^2 \;+\;r^2\cos^2\theta \,\m{d}\psi^2,
\end{flalign}
where
\begin{eqnarray}\label{E}
\rho^2& = & r^2\;+\;a^2\cos^2\theta, \nonumber\\
\Delta_r & = & \left(r^2+a^2\right)\left(1 + {r^2\over L^2}\right) - 2M,\nonumber\\
\Delta_{\theta}& = & 1 - {a^2\over L^2} \, \cos^2\theta, \nonumber\\
\Xi & = & 1 - {a^2\over L^2}.
\end{eqnarray}
Here $L$ is the asymptotic curvature length scale as above, $M$ and $a$ are strictly geometric parameters (with no direct physical meaning), and the angular coordinates are Hopf coordinates on the three-sphere\footnote{Notice that the expression for $\Delta_r$ involves $-2M$ in five dimensions, not $-2Mr$, because the ``gravitational potential'' in five dimensions is a multiple of $1/r^2$.}. The horizon ``rotates'' in the direction of $\phi$. We take it that $0 \leq a < L.$

The interior structure of this black hole is much simpler than that of a four-dimensional rotating black hole. There is of course a curvature singularity at $\rho = 0,$ but there is no inner horizon; thus there is no timelike Killing vector anywhere in the interior, which is therefore completely dynamic.

The entropy of this black hole is then
\begin{equation}\label{F}
\mathcal{S}\; =\; {\pi^2k_{\textsf{B}}c^3\left(r_{\textsf{H}}^2 + a^2\right)r_{\textsf{H}}\over 2G_5 \hbar\,\Xi}.
\end{equation}
Notice that, as discussed earlier, the entropy does indeed depend explicitly on the Newton constant $G_5$.

There has been a long debate as to the correct expression for the physical mass of this black hole. We will take the view that this question has been settled by the results of \cite{kn:gaogao}; see \cite{kn:111,kn:newgaogao} for discussions. In particular, the expression proposed in \cite{kn:gaogao} is the one that produces a physically reasonable expression for the specific angular momentum of the black hole: see below.

Following \cite{kn:gaogao}, then, we have

\begin{equation}\label{G}
\mathcal{M}\;=\;{\pi M c^2\left(3 - {a^2\over L^2}\right)\over 4\,G_5\,\left(1 - {a^2\over L^2}\right)^{3/2}}
\end{equation}
as the physical mass of the black hole.

The specific entropy of the black hole, its entropy per unit physical mass, is then
\begin{equation}\label{H}
\mathfrak{s}\;=\;{4 \pi k_{\textsf{B}} c r_{\textsf{H}}\left(1 - {a^2\over L^2}\right)^{1/2}\over \hbar \left(3 - {a^2\over L^2}\right)\left(1 + {r_{\textsf{H}}^2\over L^2}\right)}.
\end{equation}
This does not depend explicitly on $G_5$.

The angular momentum of the black hole is
\begin{equation}\label{I}
\mathcal{J}\;=\;{\pi M c^3 a\over 2\,G_5\,\left(1 - {a^2\over L^2}\right)^2},
\end{equation}
so the specific angular momentum, $\mathfrak{j}$, is
\begin{equation}\label{J}
\mathfrak{j}\;=\;{2 a c \over \left(3 - {a^2\over L^2}\right)\left(1 - {a^2\over L^2}\right)^{1/2}},
\end{equation}
which, again, does not depend on $G_5$.

\begin{figure}[!h]
\centering
\includegraphics[width=0.70\textwidth]{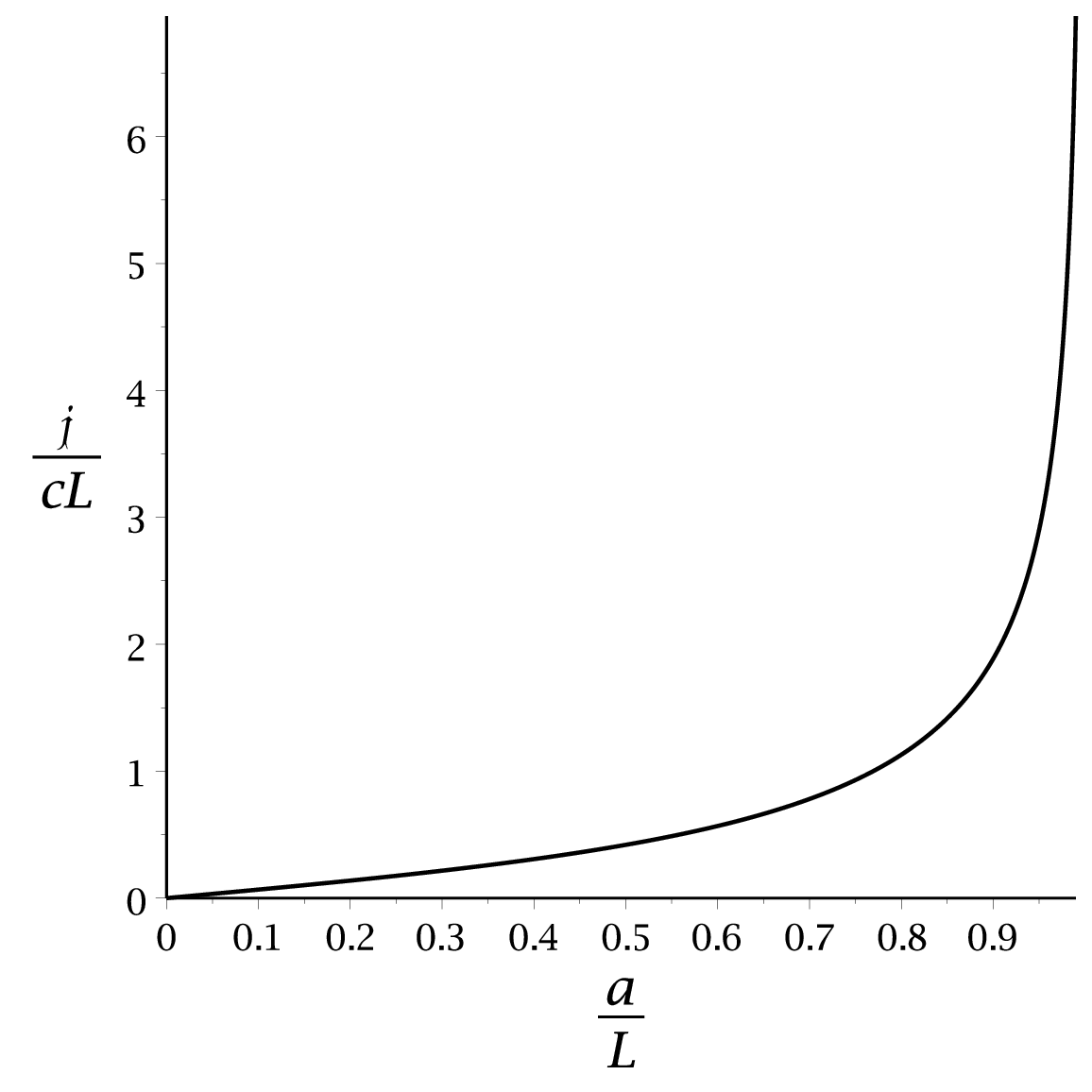}
\caption{The dimensionless specific angular momentum, $\mathfrak{j}/(cL),$ as a function of the geometric parameter $a/L$.}
\end{figure}

It is clear (see Figure 2) that $\mathfrak{j}$ increases monotonically from $0$ to infinity as $a$ ranges from $0$ to (nearly) $L$. Thus, the restriction on $a$, that it should lie between $0$ and $L$, is not really a restriction from a physical point of view, since $\mathfrak{j}$ is allowed to take \emph{any} value. (This is a result of adopting the definition of physical mass given in \cite{kn:gaogao}; it provides in fact a strong argument in favour of that definition.) Henceforth we regard $a$ as a useful formal parameter which is however just a proxy for $\mathfrak{j}.$

Note that formula (\ref{J}) gives an expression for the specific angular momentum of \emph{any} black hole of this kind. Later, when (following \cite{kn:grey}) we argue that, due to superradiance, some of the black hole angular momentum is lost to the ``galactic disc'' of a grey galaxy, the formula will be evaluated again, separately, and again it only describes the specific angular momentum of the surviving black hole, \emph{not} that of the whole system. (In particular, note that we do not use conservation of angular momentum in this work, though one could do so if one wishes to compute how much angular momentum has been transferred to the disc.)

The Hawking temperature of this black hole (see \cite{kn:111} for a discussion) is
\begin{equation}\label{K}
T\;=\;{\hbar c r_{\textsf{H}}\over 2 \pi k_{\textsf{B}}\left(1 - {a^2\over L^2}\right)^{1/2}}\left[{1 + {r_{\textsf{H}}^2\over L^2}\over r_{\textsf{H}}^2 + a^2} \,+\, {1\over L^2}\right].
\end{equation}
For reasons explained in the preceding Section, in the AdS$_5$-Schwarzschild case we are forced to confine attention to Hawking temperatures greater than or equal to $T_0 = \sqrt{2}\hbar c / \left(\pi k_{\textsf{B}}L\right)$. In the rotating case, that is no longer so: there always exist solutions for $r_{\textsf{H}}$ for any temperature. However, as we will explain in the next Section, solutions for $r_{\textsf{H}}$ when $T < T_0$ are not acceptable physically, so we continue to impose the condition $T \geq T_0$. Under that condition, there can be several solutions for $r_{\textsf{H}}$ for given values of $T$ and $a,$ but only the largest solution corresponds to a black hole with positive specific heat (see \cite{kn:111}) which can therefore reach equilibrium with its own Hawking radiation. Henceforth, then, we confine attention to these ``large'' black holes (meaning, for example, that we use this largest value of $r_{\textsf{H}}$ to compute the area of the event horizon, used in equation (\ref{F}) above).

Solving equation (\ref{K}) for $r_{\textsf{H}}$, we regard it as a known function of $T$ and $a$ (therefore, of $T$ and $\mathfrak{j}$). If we fix the temperature at some value greater than or equal to $T_0$, and thus regard $r_{\textsf{H}}$ as a function of $\mathfrak{j},$ then we find that this function has a very remarkable property: for large values of $\mathfrak{j}$ it can be made \emph{arbitrarily small} (see the upper curve in Figure 4, below). This will be crucial in our later discussion of the superradiant instability of these black holes.

If we regard $r_{\textsf{H}}$ as a function of $T$ and $\mathfrak{j},$ then equation (\ref{H}) gives the specific entropy $\mathfrak{s}$ also as a function of $T$ and $\mathfrak{j}$, and so finally we have our fundamental quantity $\mathfrak{s}\,T/\hbar$ as such a function.

A remarkable feature of this function is that it is subject to a universal upper bound: for all values of the temperature and specific angular momentum, we have \cite{kn:111}
\begin{equation}\label{LL}
\mathfrak{s}\,T/\hbar \; < \; {2 c^2\over \hbar} \; \approx \; 1.70 \times 10^{51}/(\m{kg}\cdot \m{s}).
\end{equation}
There is a suggestive way of stating this result, as follows.

When the specific angular momentum is very large, one can make a case for claiming that it should appear in the thermodynamic Euler relation (as indeed the baryonic chemical potential would, if it were not negligible in the cases we consider here). This is argued very clearly in \cite{kn:feisun,kn:jiahao}. We can incorporate this into our earlier discussion by simply defining an effective pressure $p_{\textsf{eff}}$, which by definition incorporates the usual pressure together with the effects of vorticity.

This quantity need not be subject to the usual arguments we used earlier, which put $1/3$ as the upper bound on $p/(\varrho c^2),$ but we might still require that it should satisfy an \emph{energy condition} \cite{kn:curiel} such as the Dominant Energy Condition (DEC), which requires the energy density to be positive while forbidding (the absolute value of) the pressure to exceed the energy density. Now if we write, instead of equation (\ref{A}) above,
\begin{equation}\label{LLL}
\mathfrak{s}\,T/\hbar\;=\; \left(1 + {p_{\textsf{eff}}\over \varrho c^2}\right){c^2\over \hbar},
\end{equation}
then we see that our upper bound in (\ref{LL}) (together with the obvious fact that $\mathfrak{s}\,T/\hbar$ cannot be negative) can be stated as asserting that \emph{the effective pressure satisfies the DEC}.

We also see that values of $\mathfrak{s}\,T/\hbar$ below unity correspond to a \emph{negative} effective pressure. While this is not forbidden by the DEC, it does suggest that we are in a regime where stability might be open to question.

According to our model, then, $2c^2/\hbar$ corresponds to the highest possible rate at which the quantum state evolution of strongly coupled matter can proceed. The question we want to ask now is: what is the lowest possible rate?

From this point onwards, we will typically fix the temperature at some value greater than or equal to $T_0$ and study $\mathfrak{s}\,T/\hbar$ as a function of $\mathfrak{j}.$ Unfortunately, this function is extremely complex. It is more convenient to regard $\mathfrak{s}\,T/\hbar$, with fixed temperature, as a function of $a$; this works because $a$ is a monotonically increasing function of $\mathfrak{j}$, though one must bear in mind the very different ranges of these parameters: large $\mathfrak{j}$ means that $a/L$ is close to unity.

As before, we let $\Gamma$ denote the dimensionless version of $\mathfrak{s}\,T/\hbar$ (that is, $\Gamma$ is $\mathfrak{s}\,T/\hbar$ as a multiple of ${c^2\over \hbar}$). Then using equation (\ref{H}) we have

\begin{equation}\label{L}
\Gamma\left(T^*L,\;\mathfrak{j}/(cL)\right) \;=\; {4 \pi T^* r_{\textsf{H}}\left(1 - {a^2\over L^2}\right)^{1/2}\over \left(3 - {a^2\over L^2}\right)\left(1 + {r_{\textsf{H}}^2\over L^2}\right)},
\end{equation}
where $T^* = k_{\textsf{B}}T/(\hbar c)$ as before, so that $T^*L$ is dimensionless. We saw earlier that, in current experiments, $T^*L$ ranges up to about $1.35.$ Let us set $T^*L = 2$ for the most extreme collisions foreseeable. Then we obtain Figure 3.

\begin{figure}[!h]
\centering
\includegraphics[width=0.70\textwidth]{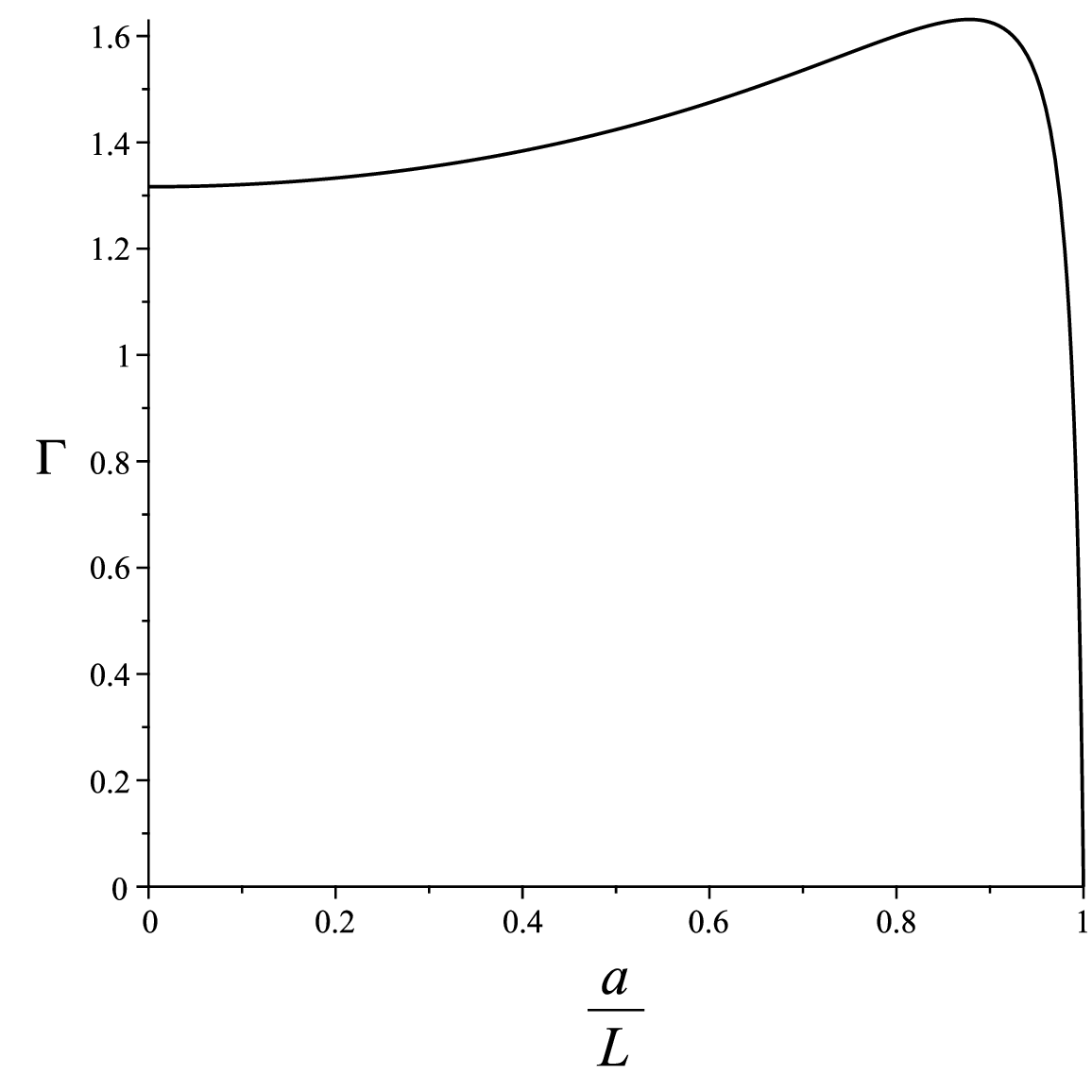}
\caption{$\Gamma\left(T^*L,\;\mathfrak{j}/(cL)\right)$ for AdS$_5$-Kerr, as a function of $a/L$ when $T^*L = 2.$}
\end{figure}

We see that the effect of increasing $a/L$ from zero (where of course it coincides with the value seen in Figure 1 when $T^*L = 2$) is to cause $\Gamma$ to increase; and in fact there is some preliminary phenomenological evidence in favour of this claim \cite{kn:kshitish}. It eventually reaches values beyond $4/3$: that is, values \emph{which are impossible in the non-rotating case}. This is the first prediction of the holographic model: values distinctly higher than $4/3$ should be observed in sufficiently vortical strongly coupled matter.

According to the model, then, increasing the vorticity \emph{at first} has much the same effect as increasing the temperature: it causes the rate of quantum state evolution to \emph{increase}. But when $a/L$ reaches about 0.9 (corresponding to $\mathfrak{j} \approx 1.89\,cL$) then any further increase in the vorticity causes $\Gamma$ to \emph{decrease}. Indeed, values \emph{smaller} than what is possible in the non-vortical case can occur (see again Figure 1). This is another predicted characteristic of extremely vortical plasmas.

This is already rather strange. Far stranger, however, is the fact, visible in Figure 3, that $\Gamma$, evaluated at some fixed temperature, can be reduced from its maximum to \emph{any} smaller value, no matter how small, by further increases in the vorticity; which means that, in principle, the rate of evolution of the quantum state on the boundary can be effectively reduced to zero. This is the bizarre conclusion to which we objected earlier.

As mentioned above, plasmas undergoing ``extremely fast rotation'', with vorticity so high that $c/\omega$ is smaller than the mean free path in the plasma, have been considered in the literature \cite{kn:buzztuch}, and such plasmas must be expected to have very unusual structures and properties, including even a spontaneous breakup into confined and deconfined zones \cite{kn:chern3}. In particular, we argued in \cite{kn:111} that a reduction in the entropy density under these extreme conditions might be expected on general grounds, due to a constriction of the relevant phase space; and this might help to explain the fact that $\Gamma$ decreases for sufficiently large specific angular momenta. However, a reduction of $\Gamma$ to almost zero seems unphysical (though, admittedly, this still marginally satisfies the DEC, as seen in equation (\ref{LLL})).

It is known \cite{kn:hawkreall} that extremal AdS$_5$-Kerr black holes are \emph{always} unstable to the phenomenon of superradiance. Although no black hole considered in this work is extremal, this observation suggests that we proceed as follows.

\addtocounter{section}{1}
\section* {\large{\textsf{5. Superradiance}}}
A straightforward calculation (see for example \cite{kn:sam}) using the First and Second Laws of (black hole) thermodynamics shows that, if a black hole rotates with an angular velocity (see below for the definition) beyond a certain critical value $\Omega_{\textsf{H}},$ then its mass must decrease: the black hole sheds some mass and angular momentum. This is an example of superradiance \cite{kn:super}.

What is much less clear is the final state of a black hole with initial angular velocity beyond $\Omega_{\textsf{H}}$, especially in the asymptotically AdS case. It is in fact not clear that the final state will even be another asymptotically AdS-Kerr black hole (see for example \cite{kn:pau} for a discussion of other possibilities).

As we discussed earlier, however, it has recently been argued convincingly \cite{kn:grey} that the final state is indeed another AdS black hole, rotating at angular velocity $\Omega_{\textsf{H}},$ surrounded by a disc of rotating matter which is coupled only weakly to the black hole \cite{kn:tidal}. Thus, any attempt to push the angular velocity of the black hole beyond $\Omega_{\textsf{H}}$ will trigger an instability, resulting in a system (a ``grey galaxy'') containing an AdS-Kerr black hole rotating at angular velocity $\Omega_{\textsf{H}}$, plus another system, to which the excess angular momentum has been transferred. This ancillary system has no interpretation in terms of a deconfined plasma, and the authors of \cite{kn:grey} interpret it as confined matter. With this result, AdS black hole superradiance can be used as a tool.

Let us be clear about what we hope to do with this tool; the situation is far from being as simple as it may appear. We will show that, as might be expected, superradiance prevents the black hole parameter $a/L$ from becoming arbitrarily close to unity \emph{when the temperature is fixed}\footnote{As stressed earlier, this leads to a bound on the specific angular momentum of the black hole, \emph{not} of the entire system.}. But it turns out that the upper bound on $a/L$ increases with the temperature; and so, by letting the temperature increase, we can drive $a/L$ arbitrarily close to unity without triggering superradiance. What we really want, however, is to show that $\Gamma$ cannot come arbitrarily close to zero, no matter what the temperature may be. This is a much more formidable task than merely keeping $a/L$ away from unity at \emph{fixed} temperature, and, in fact, it seems at first sight unlikely that this can be true.

Let us begin by briefly clarifying the definition of the ``angular velocity of the event horizon''.

We are interested in the intrinsically rotational property of the spacetime, not in what might be called the ``peculiar rotation'' of particles in orbits around the hole. We therefore focus on massless\footnote{Because we want one of them to be on the horizon.} particles with \emph{zero angular momentum}. Such particles still have non-zero angular velocities: this is \emph{frame dragging}.

Such particles on the equator (for any radial position between the horizon and infinity) have orbits satisfying
\begin{equation}\label{MM}
{ac\over r^2 \Xi} \left(\Delta_r - \left[r^2 + a^2\right]\right)\dot{t}\;+\;{1\over r^2 \Xi^2}\left(-\, a^2 \Delta_r + \left[r^2 + a^2\right]^2\right)\dot{\phi}\;=\;0,
\end{equation}
where the dot denotes a canonical affine parameter. If we picture two such particles, one on the equator of the horizon, and one on the equator at AdS conformal infinity (where frame-dragging still exists), then the \emph{difference} between the angular velocities of these two particles measures the intrinsic rotation of the horizon, and this frame-independent quantity is what is meant by the (slightly misleading) expression, ``angular velocity of the horizon'', $\omega_{\textsf{H}}$. We use this to model the vorticity of the boundary matter holographically, just as we use other quantities associated with the existence of a horizon to model the temperature, specific entropy, and so on.

This is the quantity that allows us to detect the onset of superradiance. A careful calculation shows that it is given, for the geometry represented by equations (\ref{D}) and (\ref{E}), by
\begin{equation}\label{M}
\omega_{\textsf{H}}\;=\;{ac\,\left(1 + {r_H^2\over L^2}\right)\over r_H^2 + a^2}.
\end{equation}
Hawking and Reall \cite{kn:hawkreall} found an exceedingly simple expression for the \emph{critical} angular velocity\footnote{In \cite{kn:grey}, the conventional identification of the physical mass is used, not the corrected version \cite{kn:gaogao,kn:111,kn:newgaogao}. This does not affect the computation of the critical angular velocity, which is the only explicit numerical quantity used both in \cite{kn:grey} and here, because neither definition of physical mass enters into that computation. Nor does this distinction affect the qualitative picture of the end state advanced in \cite{kn:grey}. It certainly \emph{will} affect the numerical details, for example of the distribution of energy and angular momentum between the core and the disc of the grey galaxy, and this merits further investigation (perhaps using the methods of \cite{kn:tidal}).}, the value beyond which superradiance is triggered:
\begin{equation}\label{N}
\Omega_{\textsf{H}}\;=\;c/L.
\end{equation}
Notice that $L$ has no interpretation as a ``radius'' for the horizon (or for the sphere at infinity), so, contrary to what is sometimes said, this relation has nothing to do with causality.

If we accept the suggestion in Section 2 above that $T_0$ (see equation (\ref{B})) is approximated by the QGP crossover temperature (around 150 MeV/$k_{\textsf{B}}$), then $c/\Omega_{\textsf{H}}$ is well under one femtometre, and so the onset of superradiance in the bulk corresponds to the ``extremely fast'' domain of angular velocities discussed in \cite{kn:buzztuch}, which we mentioned earlier. This means that our holographic picture of the situation is internally consistent, in the sense that we expected ``extremely fast rotation'' to correspond to small specific entropies and therefore to the region near to $a/L = 1$ in Figure 3.

Substituting (\ref{N}) into equation (\ref{M}), we obtain a relation between $r_{\textsf{H}}$ and $a/L$ in the critical case. The graph of this function is shown as the lower curve in Figure 4; superradiance occurs if $r_{\textsf{H}}$ falls \emph{below} that curve. Bear in mind that $a/L$ is a certain monotonically increasing function of $\mathfrak{j}/(cL)$ here.

When $T \geq T_0$ (which means that $T^*L \geq \sqrt{2}/\pi$), the graph of $r_{\textsf{H}}$ as a function of $a/L$ always resembles the upper curve shown in Figure 4 (which portrays the graph when $T^*L = 2,$ well above $\sqrt{2}/\pi \approx 0.450$). That is, near $a/L = 0$ it lies above the curve we have just been discussing, meaning that there is of course no superradiance for low values of $a/L$. Eventually, however, it (always) bends down to arbitrarily small values, meaning that it \emph{must} intersect the lower curve: the black hole becomes superradiant at sufficiently large values of $\mathfrak{j}$, in fact for any value greater than about $6.13\,cL$ in this example. (The lower curve never changes, but the upper one does, in response to changes in $T$; therefore, the value of $a/L$ at the intersection is a function of $T$. See the next Section for more detail.)

\begin{figure}[!h]
\centering
\includegraphics[width=0.80\textwidth]{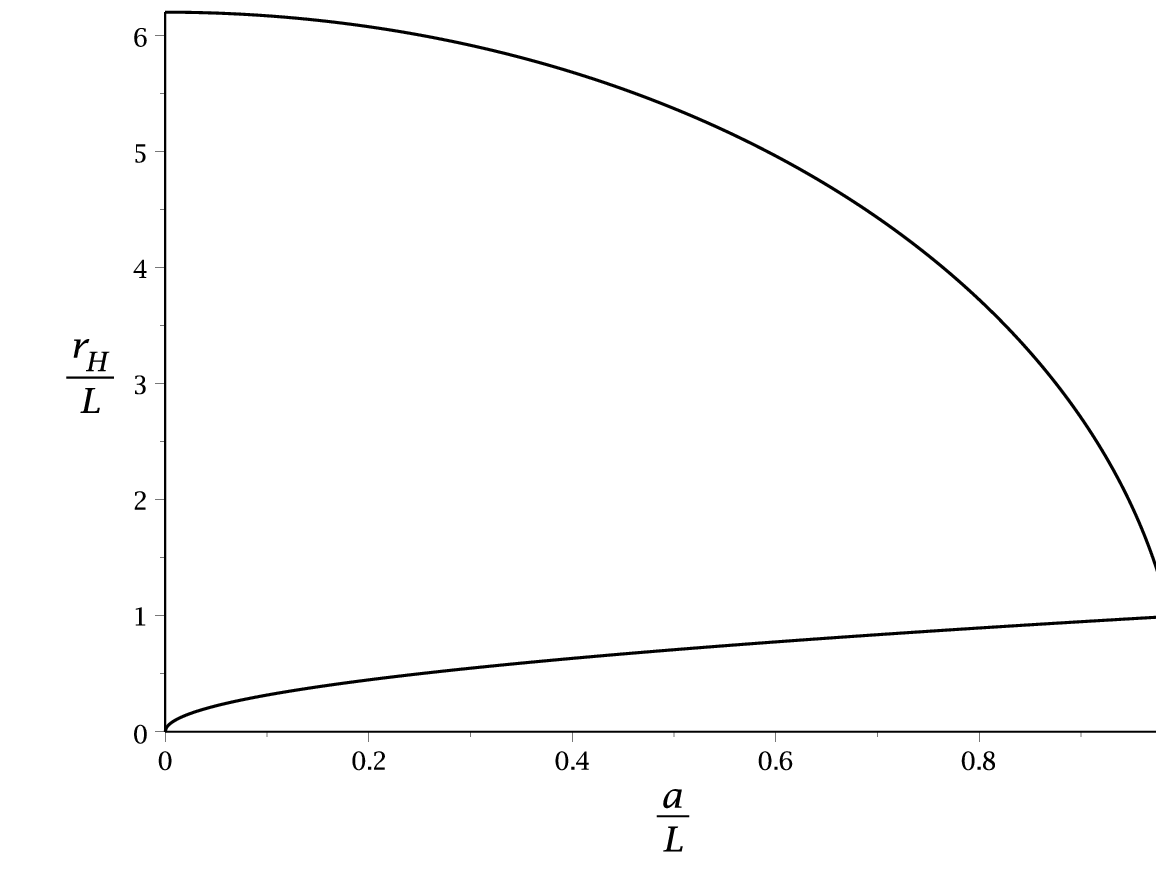}
\caption{Upper curve: the radial coordinate at the horizon, for $T^*L = 2,$ as a function of $a/L.$ Lower curve: the black hole is unstable against superradiance for any value of $r_{\textsf{H}}/L$ below this curve. The intersection in this case is at $a/L \approx 0.987,$ corresponding to $\mathfrak{j}/(cL) \approx 6.13.$ }
\end{figure}

However, it is important to understand that the fact that the two curves always intersect was by no means foreordained: it only happened because $r_{\textsf{H}}$ falls to arbitrarily small values when $a/L$ nears unity. (In fact, the intersection occurs here because we have adopted the definition of physical mass proposed in \cite{kn:gaogao}; with the previously generally accepted definition, the curves never intersect and superradiance imposes no restrictions. See \cite{kn:109}.)

We mentioned earlier that the rotating case differs from the AdS$_5$-Schwarzschild case: here $r_{\textsf{H}}$ always exists for all values of $T$ and $a$. But when $T < T_0$, the graph changes discontinuously: it suddenly assumes the form shown in Figure 5 as the lower curve (where we have taken $T^*L = 0.35)$. The upper curve is the curve below which superradiance occurs, so we see in this case that the black hole is unstable for any value of the angular momentum, \emph{no matter how small} it may be\footnote{In this particular example, the black hole is unstable for all values of $a/L$. For some values of $T^*L$ closer to $\sqrt{2}/\pi,$ however, the lower curve can protrude above the upper one for a narrow range of values of $a/L$ away from zero; but it is always true that these black holes are apparently unstable to superradiance at arbitrarily small values of $a/L$.}. This does not make sense physically ---$\,$ it would be absurd to have superradiance triggered by an arbitrarily small specific angular momentum ---$\,$ so we continue to assume that temperatures below $T_0$ do not occur in strongly coupled matter\footnote{In this model, the critical temperature is independent of the specific angular momentum. This is however due to the simplicity of our model. It has been known for some time \cite{kn:mcinnes} that, in more elaborate holographic models, angular momentum can in fact have some effect on the predicted minimal possible temperature: see \cite{kn:yidian} for a comprehensive discussion of this.}.

\begin{figure}[!h]
\centering
\includegraphics[width=0.70\textwidth]{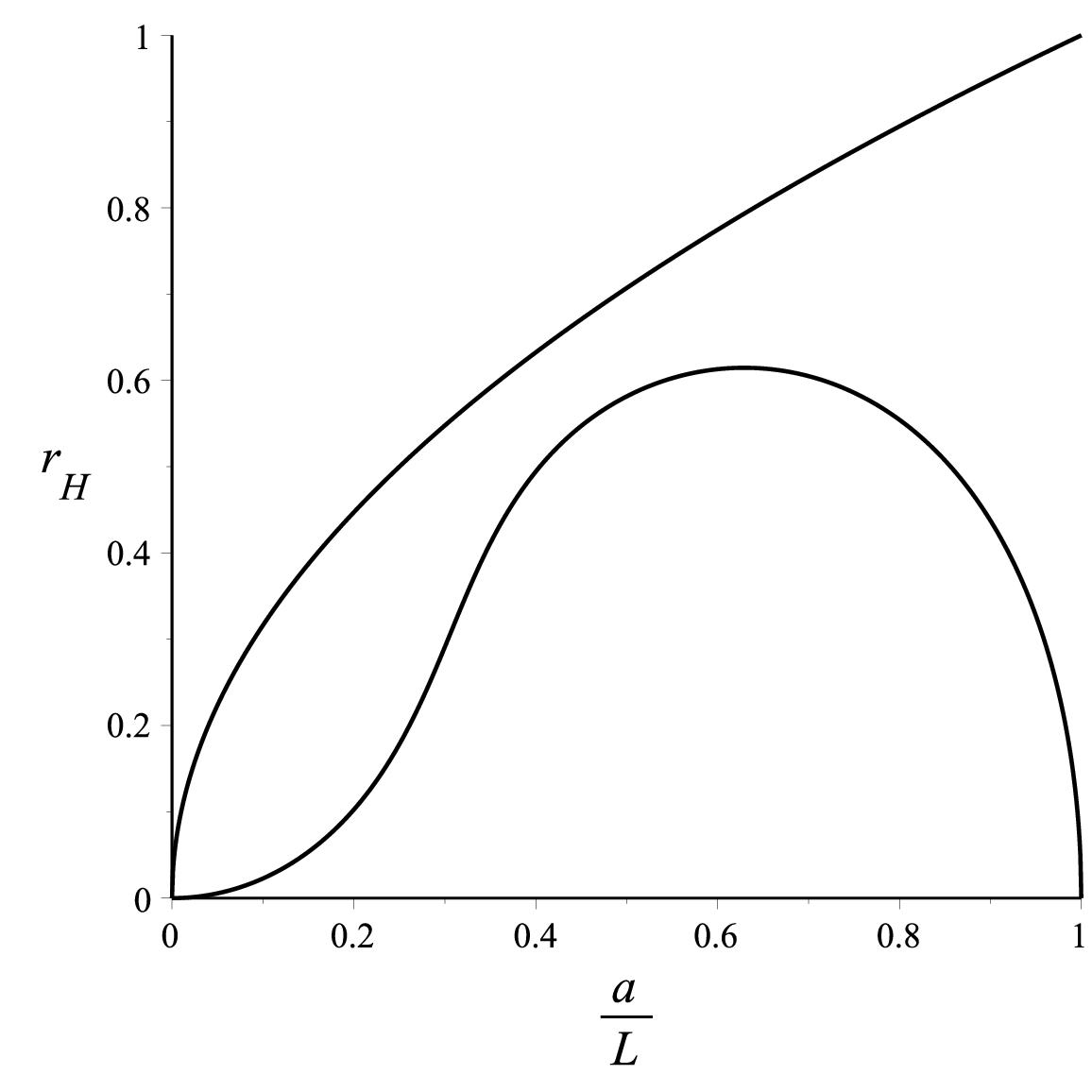}
\caption{Lower curve: the radial coordinate at the horizon, for $T^*L = 0.35 < \sqrt{2}/\pi,$ as a function of $a/L.$ Upper curve: the black hole is unstable against superradiance for any value of $r_{\textsf{H}}$ below this curve.}
\end{figure}

In summary, then, we see that $a/L$ cannot come arbitrarily close to unity without triggering an instability: the critical value of $a/L$ for a given temperature $T$ above $T_0$ can be computed by finding the intersection point, as shown in an example in Figure 4. According to the grey galaxy hypothesis, if the black hole is given an angular momentum that takes $a/L$ beyond this critical value, it will lose that angular momentum to the disc, and revert back to the critical angular velocity, $\Omega_{\textsf{H}}\;=\;c/L.$ In effect, then, we have an upper bound on $a/L$ and therefore on the specific angular momentum.

We repeat however that \emph{these considerations alone do not solve our problem}. What we have shown is that, for each fixed temperature, there is a maximal value of $a/L$, $a_{\textsf{max}},$ which is strictly smaller than unity. But in fact, as the temperature increases, the intersection point slides to the right, and $a/L$ \emph{tends} to unity. In view of Figure 3, this \emph{seems} to mean that $\Gamma_{\textsf{min}}$, the minimal value of $\Gamma$ at a given temperature, can be made arbitrarily small by taking the temperature sufficiently large, and this is not what we want.

Bearing this danger in mind, let us proceed to compute $\Gamma_{\textsf{min}}$.

\addtocounter{section}{1}
\section* {\large{\textsf{6. A Lower Bound on the Dimensionless Rate}}}
We have just seen that $a_{\textsf{max}}$ is found by solving for the intersection point seen, in an example, in Figure 4. As we mentioned, the location of the intersection is a function of $T$ or $T^*$: one finds $a_{\textsf{max}}$ by choosing the larger root after solving (for $a$ in terms of $T^*$) the equation
\begin{equation}\label{O}
2\pi T^*L \;=\; {1\over \sqrt{a_{\textsf{max}}/L}}\,\sqrt{{1 + a_{\textsf{max}}/L \over 1 - a_{\textsf{max}}/L}}.
\end{equation}
Similarly the value of $r_{\textsf{H}}$ at the intersection can be expressed as a function of $T^*$. Substitution into equation (\ref{L}) then gives us the value of $\Gamma$ at the intersection, as a function of $T^*$.

The function expressing the value of $\Gamma$ at $a/L = a_{\textsf{max}}$  is shown in Figure 6. (In Figures 6 and 7, the range of $T^*L$ begins at $\sqrt{2}/\pi,$ for reasons explained earlier.)

\begin{figure}[!h]
\centering
\includegraphics[width=0.70\textwidth]{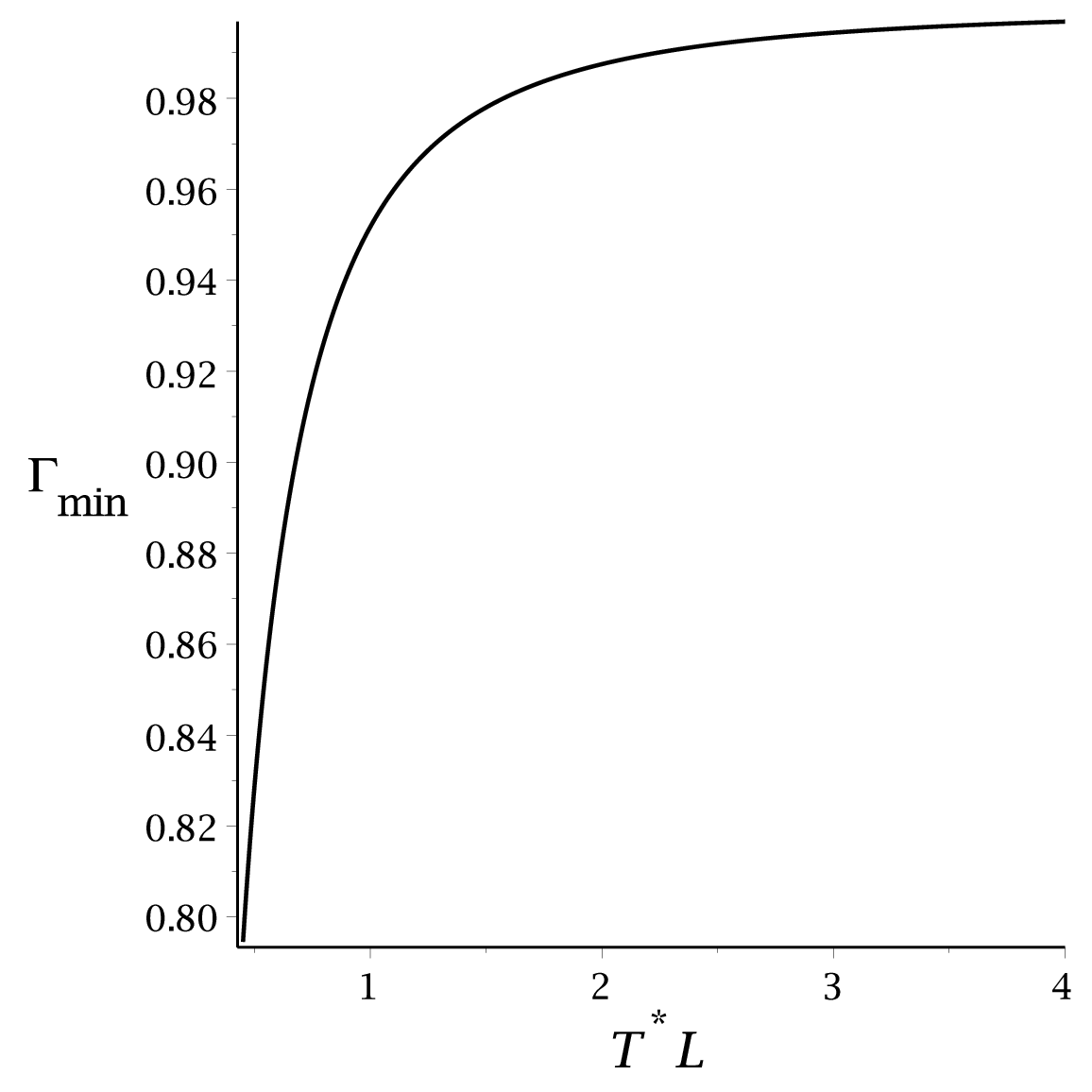}
\caption{$\Gamma$ at $a = a_{\textsf{max}},$ as a function of temperature. }
\end{figure}

We claim that the function with graph as shown gives the minimal possible value of $\Gamma,$ as a function of temperature, taking superradiance into account. To see this, we simply note that, from Figure 3, this minimum must occur either at $a = a_{\textsf{max}}$ or at $a = 0.$ But in fact the latter value is always larger, as can be seen in Figure 7. Thus the graph in Figure 6 does indeed give the minimal $\Gamma$ as a function of $T^*L$.

\begin{figure}[!h]
\centering
\includegraphics[width=0.70\textwidth]{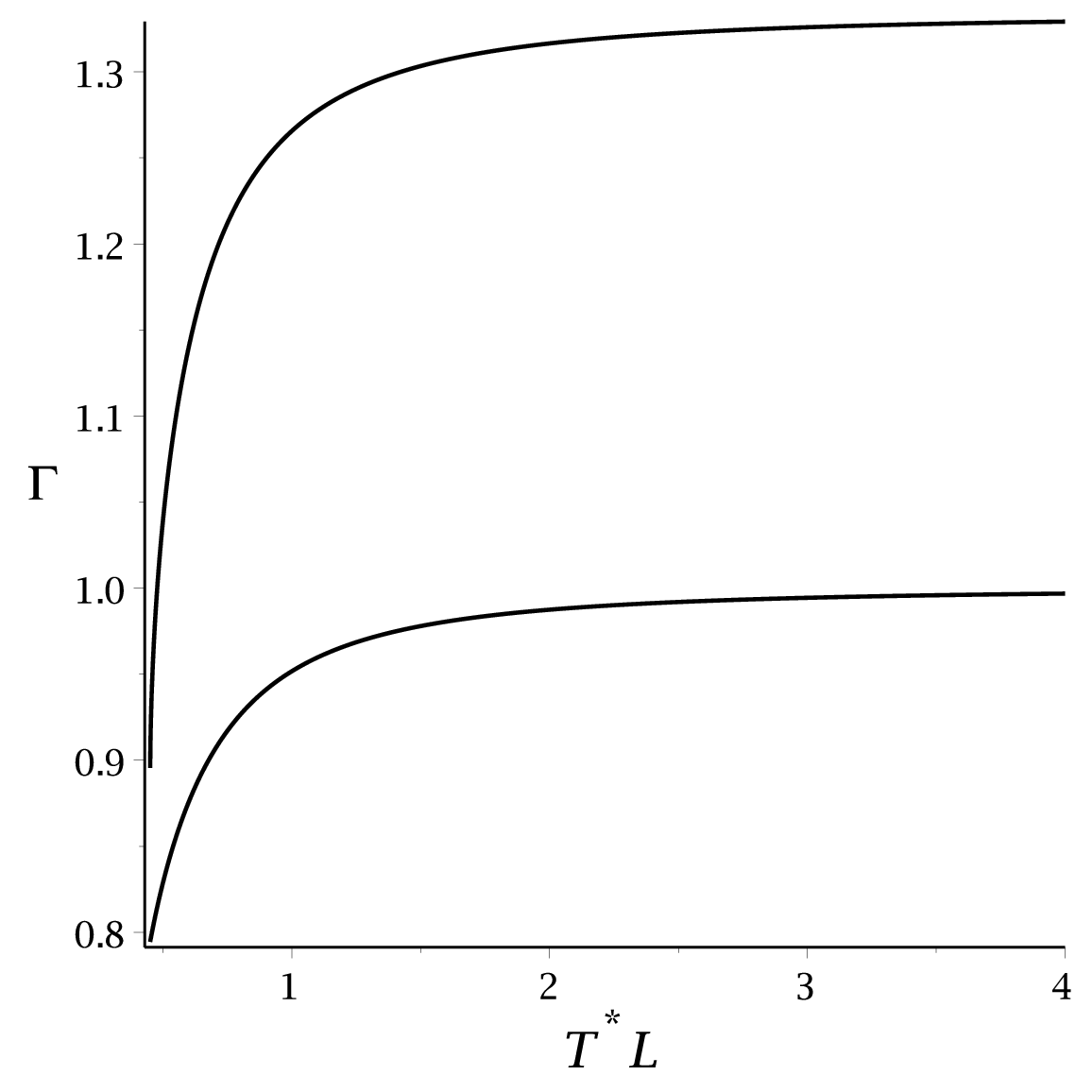}
\caption{$\Gamma_{\textsf{min}}$ (lower graph) vs $\Gamma$ at a = 0 (upper graph). Here $T^* = k_{\textsf{B}}T/(\hbar c)$ as usual.}
\end{figure}

We see that, as $T$ becomes large, $\Gamma_{\textsf{min}}$ \emph{does not} fall to small values; in fact, it actually increases. This is because the graph in Figure 3 is extremely steep near $a/L = 1$, and, as $T$ increases, it becomes ever more steep, as the maximum is pushed to the right; this delicate interplay allows $\Gamma_{\textsf{min}}$ to remain relatively large even as $a_{\textsf{max}}/L$ tends to unity. Thus for example, as we saw earlier, when $T^*L = 2$, we have $a_{\textsf{max}}/L \approx 0.987,$ and yet $\Gamma_{\textsf{min}} \approx 0.987$ there\footnote{The (approximate) similarity of these two numbers is merely coincidental.}.

In fact, one can prove analytically that $\Gamma_{\textsf{min}}$ is always less than unity, that it increases monotonically as a function of $T^*L$, and that it has an upper bound given by
\begin{equation}\label{P}
\lim_{T \rightarrow \infty} \Gamma_{\textsf{min}} \;=\; 1.
\end{equation}
Actually, for all but the very lowest temperatures, $\Gamma_{\textsf{min}}$ is quite close to unity: for example, at $T^*L = 1$, it is approximately $0.952$. Its smallest value, occurring of course at $T = T_0$, the minimal possible temperature of strongly coupled matter in our model, is approximately $0.795$.

In terms of the ``effective pressure'' (equation (\ref{LLL})): in the absence of superradiance, large specific angular momenta can drive $p_{\textsf{eff}}/(\varrho c^2)$ down to the smallest value compatible with the Dominant Energy Condition, that is, close to $- 1$. Superradiance prevents this: $p_{\textsf{eff}}/(\varrho c^2)$ can still be driven down to negative values (though for all practical purposes this only really happens at the very lowest temperatures), but these values are small in absolute value. In other words, for all but the very lowest temperatures, superradiance tends to suppress negative effective pressures. It would be interesting to have a better understanding of the physics of this.

The reader may find it instructive to compare Figure 6 with Figure 1. According to the holographic model, the effect of vorticity is to slow down the quantum state evolution on the boundary: but the extent of the slowing is not very great.

To summarise, then: superradiance ensures that all stable AdS$_5$-Kerr black holes, no matter what their specific angular momenta and temperatures may be, satisfy a universal counterpart to the inequality (\ref{LL}):
\begin{equation}\label{Q}
\mathfrak{s}\,T/\hbar \; \geq \approx \; 0.795 \; {c^2\over \hbar} \; \approx \; 6.77 \times 10^{50}/(\m{kg}\cdot \m{s}).
\end{equation}
This is still many orders of magnitude larger than the value for normal matter. Even in the most extreme circumstances, the holographic model predicts that the rate of quantum state evolution of strongly coupled matter is always large. The dual statement is that the interior of a rapidly rotating AdS black hole evolves more slowly than that of an AdS Schwarzschild black hole with the same Hawking temperature, but also that this interior evolution cannot be (nearly) halted, contrary to what is suggested in \cite{kn:110}.

What we have shown is that superradiance protects quantum state evolution on the boundary. However, one would expect a system rotating at a gigantic angular velocity to be potentially unstable in a variety of ways. Can we give any evidence that superradiance is indeed the true explanation of the fact that the rate of quantum state evolution cannot in reality be arbitrarily small?

The reader will have noticed that the superradiant upper bound on $a/L$ (at given temperature) implies, through equation (\ref{J}), an upper bound on the specific angular momentum of the black hole; and this in turn, through holography, implies an upper bound on the specific angular momentum of the strongly coupled matter on the boundary. We now argue that this provides (in principle) a way of identifying superradiance as the physically relevant mechanism here.

\addtocounter{section}{1}
\section* {\large{\textsf{7. An Upper Bound on the Specific Angular Momentum}}}
We saw earlier that, given any temperature above $T_0$, we can place an upper bound on $a/L$ using superradiance. Using equation (\ref{J}) one finds that this means that there is a predicted upper bound on $\mathfrak{j}/(cL).$ It is interesting to ask how this compares with actual values of the specific angular momentum, from experimental observations on the QGP. We can think of this as a way of checking whether the theory is anywhere near making contact with the data.

The difficulty here is that we do not know the value of $L$. However, we can circumvent this if we are willing to conjecture, as discussed in Section 2, that our minimal temperature $T_0$ (see equation (\ref{B})) is approximately equal to the crossover temperature, which is around 150 MeV/$k_{\textsf{B}}$. For then we can write
\begin{equation}\label{R}
{\mathfrak{j}\over cL} \;=\; {\pi k_{\textsf{B}}\,\mathfrak{j}\, T_0 \over \sqrt{2}\hbar c^2},
\end{equation}
and this involves $T_0$, not $L$.

Now $\mathfrak{j}$ is interpreted as the ratio of the angular momentum and energy densities for plasmas produced in collisions at a given impact energy. In principle, the temperature and the two densities can be extracted from the data; and then we can compare the right side of equation (\ref{R}) with the predicted upper bound for that temperature.

Unfortunately it is in practice extremely difficult to do this. Even for the energy density in central collisions, there are endless complications, discussed in great detail in \cite{kn:phobos}; and the evaluation of the angular momentum density presents still more difficulties. We will proceed with simple assumptions and accept that the results will not be precise.

For definiteness, let us focus on the collisions at 200 GeV impact energy described in \cite{kn:STARcoll} (where the first confirmed observations of QGP vorticity were reported). There, the value given for angular momentum is an \emph{averaged} value over many collisions at varying centralities. Some individual collisions, at favourable impact parameters, may in theory \cite{kn:jiang} result in far larger angular momenta than the values given in \cite{kn:STARcoll}, but we only have observational access to this average. Therefore, we do our computations for an imagined ``typical collision'' producing a plasma with angular momentum $1000 \,\hbar$ as reported in \cite{kn:STARcoll}, at a typical impact parameter of 7 femtometres (fm), that being the approximate radius of a gold nucleus. We follow \cite{kn:phobos} and take it that the energy density immediately after equilibration is around 3000 MeV/fm$^3$.

We have to take into account that the angular momentum is deposited into the overlap volume of the two colliding ions, which can be computed using the sphere-sphere intersection formula \cite{kn:weiss}; but we also have to allow for the Lorentz contraction factor, which, immediately after equilibration, is estimated in \cite{kn:phobos} to be around 7. Our final estimate for this ``typical collision'' is that the angular momentum resides in a volume of about 64 fm$^3$.

Substituting all these quantities, with $T_0 \approx \, $ 150 MeV/$k_{\textsf{B}}$, into the right side of equation (\ref{R}), we estimate
\begin{equation}\label{S}
{\mathfrak{j}\over cL} \;\approx\; 1.7;
\end{equation}
this is (at best) an order-of-magnitude estimate, so error estimates are pointless here (and indeed none is given for the value of the angular momentum reported in \cite{kn:STARcoll}).

Taking the temperature to be, as suggested in \cite{kn:STARcoll}, about 200 MeV/$k_{\textsf{B}}$, we see from our discussion in Section 2 that we have $T^*L \approx 1.$ The analogue of the intersection in Figure 4 occurs at $a_{\textsf{max}}/L \approx 0.948,$ and this means, from equation (\ref{J}), that the maximal value of $\mathfrak{j}/(cL)$ is predicted to be
\begin{equation}\label{T}
{\mathfrak{j}_{\textsf{max}}\over cL} \;\approx\; 2.8.
\end{equation}
We do not expect the reader to attach much importance to the fact that the data are consistent with this upper bound. The key point is that the correct values of the two numbers in (\ref{S}) and (\ref{T}) may well be of about the same order of magnitude.

In short, the typical vortical systems produced in the RHIC collisions at 200 GeV impact energy could be pushing close to the upper bound imposed by superradiance, if that is indeed the relevant effect here.

The situation becomes more interesting, but unfortunately also more obscure, when we turn to collisions at higher impact energies, such as those of lead ions studied by the ALICE collaboration \cite{kn:ALICE} at the LHC. These are in the 5 TeV impact energy range, far higher than before, and naturally the temperature is considerably higher than at the RHIC, by roughly a factor of 2. The upper bound on $\mathfrak{j}/(cL)$ is a steep, approximately linear function of the temperature: see Figure 8. (We assume that the QGP is still formed at these temperatures; this is not indisputable.)

\begin{figure}[!h]
\centering
\includegraphics[width=0.70\textwidth]{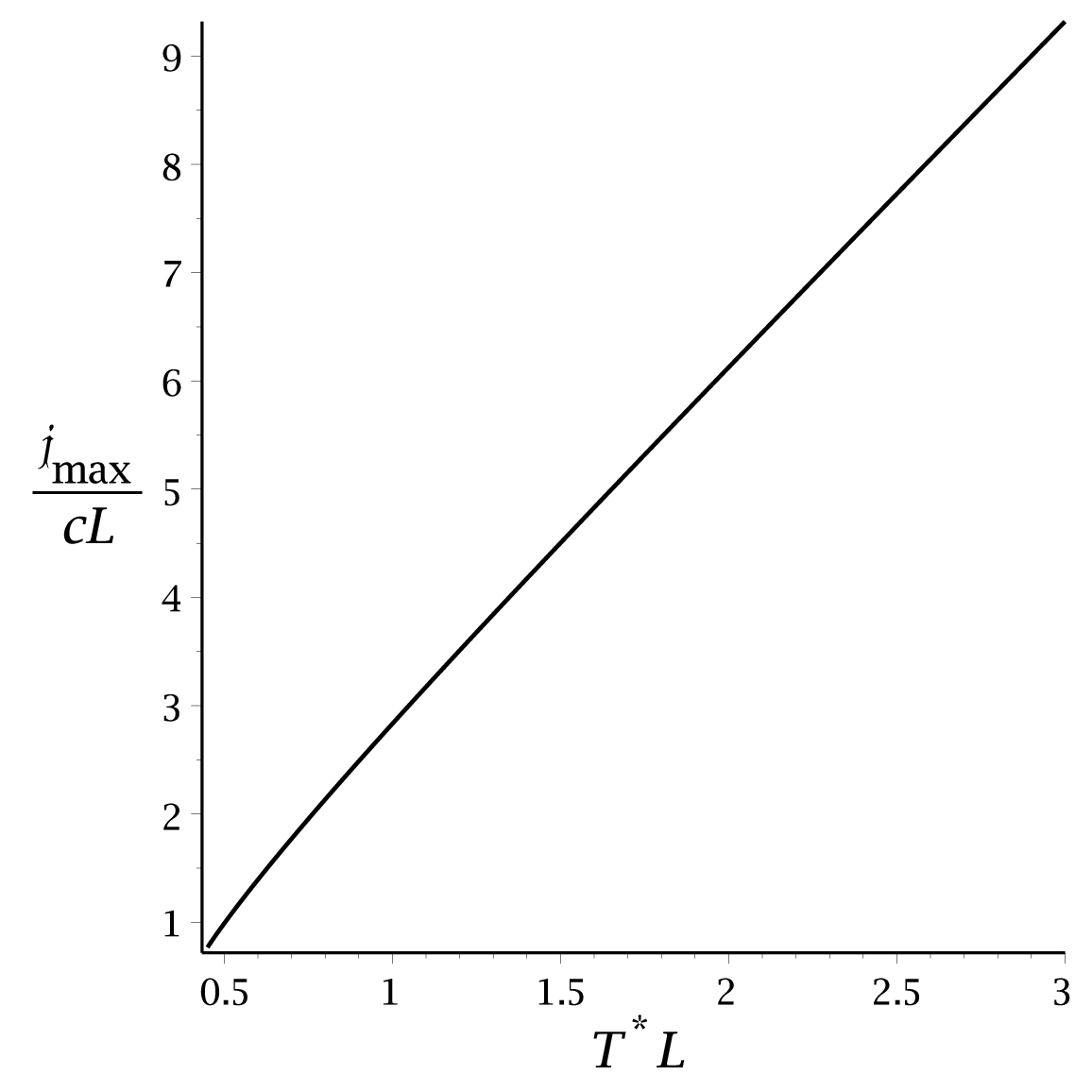}
\caption{$\mathfrak{j}_{\textsf{max}}/(cL)$ as a function of temperature.}
\end{figure}

Thus the superradiant upper bound for these collisions is around $6.1$, much higher than before.

Comparison with data is more difficult in this case, because vorticity becomes \emph{less}, not more, detectable at higher impact energies: see for example \cite{kn:becca,kn:newalice}. Thus a final decision awaits further data and analysis. Let us nevertheless proceed, on a very tentative basis.

Again, we need to estimate $\mathfrak{j},$ the ratio of the angular momentum and mass/energy densities. Both densities tend to increase with the large jump in impact energy.

The energy density is much larger, by a factor of four or more, and this tends of course to reduce $\mathfrak{j}/(cL)$. However, the amount of angular momentum imparted to the plasma at given impact parameter increases approximately linearly with the impact energy \cite{kn:jiang}, so the increase in the angular momentum density will greatly outweigh the increase in the energy density. Thus, in this case, a typical peripheral (non-central) collision will produce initial values of $\mathfrak{j}/(cL)$ well beyond the superradiant bound, $6.1$.

Granting this, what should we expect to see?

According to holographic duality, generating such values for $\mathfrak{j}/(cL)$ means that the dual bulk black hole undergoes a superradiant instability, and the hole will shed some energy and angular momentum. What the ``grey galaxy'' model \cite{kn:grey} teaches us specifically is that \emph{the end state of this process will be a black hole on the brink of superradiance}, that is, with $\mathfrak{j} = \mathfrak{j}_{\textsf{max}}$, plus a ``galactic disc'', to which the excess angular momentum has been transferred. The latter conjecturally represents confined matter, to which the duality does not apply.

If all this is correct, then, the surviving plasma will satisfy
\begin{equation}\label{U}
\mathfrak{j}_{\textsf{max}} \;\approx\; 6.1\,\times \, {\sqrt{2} \hbar c^2\over \pi k_{\textsf{B}} T_0} \; \approx \; 3 \times 10^{- 3}\; \hbar c^2\, \m{MeV}^{-1}.
\end{equation}
The corresponding angular velocity is about $1.7$ $c\,$fm$^{- 1}$ (compare with the phenomenological values explored in, for example, \cite{kn:kshitish}).

The point is that since a generic collision at these impact energies triggers superradiance in the dual theory, such a collision should produce a plasma with specific angular momentum given in equation (\ref{U}). That is, one expects to find that the vorticity in these experiments, if it can be observed more precisely than is currently possible, will be \emph{clustered} around some specific value, perhaps corresponding roughly to an angular velocity around $1.7$ $c\,$fm$^{- 1}$ and specific angular momentum about $3 \times 10^{- 3}\; \hbar c^2\, \m{MeV}^{-1}$. If this prediction is (of course, roughly) confirmed, that will be evidence that superradiance is indeed the effect underlying our conclusions above.

\addtocounter{section}{1}
\section* {\large{\textsf{8. Conclusion}}}
The ``AdS'' in AdS/CFT is nearly always a black hole spacetime, and so it is highly desirable to have a complete holographic account of these objects, including their interiors. Yet, as we have stressed, the most characteristic property of the interior, its time dependence, seems to be the hardest property to fit into a dual picture.

The suggestion that the interior time-dependence is dual to the post-equilibrium quantum state evolution of the strongly coupled matter on the boundary seems very natural. But neither process is directly observable, so how is the extended duality to be studied? Here we have proposed that the key is the dependence of the \emph{rate} of boundary matter quantum state evolution on the black hole parameters, in particular on the Hawking temperature and the specific angular momentum. We do not know much about the rate, but it seems almost certain that it cannot be (arbitrarily close to) zero. This simple observation implies the existence of some kind of bound on the specific angular momentum of strongly coupled matter, and this allows us, in principle at least, to make contact with experimental observations.

Experimental work on the ultra-vortical QGP produced in peripheral collisions of heavy ions continues apace (see for example \cite{kn:be}). Although we do not expect quantitative agreement with our very simple holographic model, even qualitative agreement (for example, clustering of vorticities around some large maximal value, not necessarily very near to our estimate here) would be of very considerable interest.

\addtocounter{section}{1}
\section*{\large{\textsf{Acknowledgement}}}
The author is grateful to Prof. Maxim Chernodub for drawing his attention to the important references \cite{kn:chern3,kn:chern1,kn:chern2,kn:chern4}, and to Dr. Soon Wanmei for useful comments.

\end{document}